\documentclass[%
 reprint,
superscriptaddress,
 amsmath,amssymb,
 aps,
]{revtex4-2}
\usepackage{physics}
\usepackage{bbold}
\usepackage{amsmath}
\usepackage{latexsym}
\usepackage{graphicx}
\usepackage{amsthm}
\usepackage{hyperref}
\usepackage{lipsum}
\usepackage[english]{babel}
\usepackage{slashed}
\usepackage[compat=1.1.0]{tikz-feynman}
\usepackage{appendix}
\usepackage{subfig}
\usepackage{multirow}
\usepackage{mathtools}
\usepackage{subfig}

\theoremstyle{plain}

\begin{document}

\title{Entanglement Entropy in Scalar Quantum Electrodynamics} 

\author{Samuel Fedida}
\affiliation{Centre for Quantum Information and Foundations, DAMTP, Centre for Mathematical Sciences, University of Cambridge, Wilberforce Road, Cambridge CB3 0WA, UK}
\author{Anupam Mazumdar}
\affiliation{Van Swinderen Institute, University of Groningen, Groningen 9747 AG, The Netherlands}
\author{Sougato Bose}
\author{Alessio Serafini}
\affiliation{Department of Physics \& Astronomy, University College London, Gower Street, London WC1E 6BT, UK}

\date{\today}

\begin{abstract}
    We find the entanglement entropy of a subregion of the vacuum state in scalar quantum electrodynamics, working perturbatively to the 2-loops level. Doing so leads us to derive the Maxwell-Proca propagator in conical Euclidean space. The area law of entanglement entropy is recovered in both the massive and massless limits of the theory, as is expected. These results yield the renormalization group flow of entanglement entropy, and we find that loop contributions suppress entanglement entropy. We highlight these results in the light of the renormalization group flow of couplings and correlators, which are increased in scalar quantum electrodynamics, so that the potential tension between the increase in correlations between two points of spacetime and the decrease in entanglement entropy between two regions of spacetime with energy is discussed. We indeed show that the vacuum of a subregion of spacetime purifies with energy in scalar quantum electrodynamics, which is related to the concept of screening.
\end{abstract}

\maketitle

\onecolumngrid

\section{Introduction}

The tools of quantum information theory have continuously brought insights into numerous seemingly unrelated areas of physics. There is notably a growing interest in the concepts of entanglement and entropy within the context of high-energy physics and phenomenology, as well as in condensed matter theory. More particularly, there have been recent studies in the generation of entanglement in ordinary local quantum field theories (QFTs) \cite{Calabrese2004,Casini:2009sr,Metlitski:2009iyg,  Hertzberg2013,Seki2015,Witten:2018zxz,Casini:2022rlv, Headrick2018TASILO,Hollands:2017dov,Fedida2023}; in the context of holographic quantum field theories, see~\cite{Casini:2011kv,
Rangamani:2016dms}; in the context of area-law of entanglement entropy~\cite{Bekenstein:1981,Sorkin:1986,Srednicki:1993im}; in scattering processes of particles~\cite{Balasubramanian:2011wt,Peschanski2016,Bose:2017nin,Peschanski2019, Marshman:2019sne, Carney_2019,Bose:2022uxe,Biswas:2022qto,Carney:2021vvt,
Elahi:2023ozf,Chakraborty:2023kel,Bose2020}; and within non-local quantum field theory (NLQFT) \cite{Landry2024,Vinckers:2023grv}. In the present paper, we focus on standard (local) QFT and consider the properties of entanglement entropy within scalar quantum electrodynamics (QED).

\bigskip

In the context of particle physics, neutrino oscillations are being actively investigated using tools from quantum information theory \cite{Blasone2009,Alok2016,Song2018,Wang2020,Bittencourt2022}, where such methods can provide numerous insights into exotic mechanisms in relativistic settings. Moreover, a systematic inquiry of the entanglement generated in $2\to2$ tree-level scatterings in spinor QED between helicity/polarisation degrees of freedom has been undertaken, both for pure initial states \cite{CerveraLierta2016} and for general (potentially mixed) states \cite{Fedida2023}, where the generation of Von Neumann entropy was also discussed. Here, we shall go beyond tree-level processes, allowing us to delve into the renormalization of such quantum information-theoretic properties and examine vacuum fluctuations rather than scattering processes. 

\bigskip

The dependence of entanglement entropy on the partition size of the system in pure Maxwell theory (a conformal field theory) has been the subject of some debate \cite{Donnelly2015,Pretko2018}. The behaviour of geometric entropy in the context of lattice gauge theories has also been explored \cite{Calabrese2004,Donnelly2012,Casini2014,Ghosh2015,Radicevic2016,Donnelly2016,Soni2016}. Interestingly, the properties of entanglement entropy in $\phi^3$ and $\phi^4$ theories have been studied by Hertzberg \cite{Hertzberg2013} up to two-loop orders, and a tentative renormalization procedure has been undertaken. Significant work has been conducted to understand the renormalization of entropy in conformal field theory \cite{Holzhey1994} and in $\phi^4$ theory \cite{Pang2021}, as well as the regularization of entropy in more general QFT settings \cite{Kudlerflam2023}. A generalised 2PI formalism has also been used to approach similar topics nonperturbatively, and the non-Gaussianity of entanglement entropy induced by the Wilson renormalization group procedure has been analysed \cite{Iso2021a,Iso2021b,Iso2021c}.

\bigskip

In this paper, we shall pursue such efforts by computing the ground state entanglement entropy in scalar QED perturbatively up to two-loop order. Doing so requires us to work in conical Euclidean space - where the so-called replica trick, detailed in Section \ref{Setting and Preliminaries} below, is used to recover the flat spacetime limit~\cite{VanRaamsdonk2017}. Since the entropy depends on the propagators of the theory of interest in conical space, we shall derive in Section \ref{Greens functions in Conical Space} the Maxwell-Proca propagators in such a setting. In Section \ref{Entanglement entropy in the free theory} we will show that the area law of entanglement entropy is indeed obtained, as can be expected, and we shall explore in Section \ref{Entanglement entropy in scalar qed} the renormalization of vacuum entropy in such a context. We will further consider the interplay between the renormalization of couplings and correlators with that of entanglement entropy.

\bigskip

\section{Setting and Preliminaries}

\label{Setting and Preliminaries}

We work in units $\hbar=1$, Feynman gauge $\xi=1$, and Euclidean space with negative signature. We consider a QFT (in our case, scalar QED), which lies on an infinitely large and flat $D=(1+d)$-dimensional spacetime $\Omega$. Going to Euclidean time through a Wick rotation, we divide $\Omega$ into two regions $\Omega = A \cup \bar{A}$ through an (arbitrary) cut on the real negative axis, such that the sub-spaces $A$ and $\bar{A}$ have a flat dividing boundary of dimension $d_\perp = d-1$. The density matrix of the ground state of the QFT on the sub-region of interest $A$ is obtained by tracing out the degrees of freedom in the region $\bar{A}$, that is, $\rho_A = \Tr_{\bar{A}}(\rho)$. The associated entanglement entropy of the correlations between vacuum fluctuations is given by the Von Neumann entropy of the reduced density matrix
\begin{equation}
    \label{VNS definition}
    S_E(\rho_A) := -\Tr(\rho_A \ln(\rho_A))
\end{equation}
We can rewrite this using the replica trick
\begin{equation}
    \label{VNS replica}
    S_E(\rho_A) = - \frac{\partial}{\partial n} \ln(\Tr(\rho_A^n))\Big|_{n\to 1}
\end{equation}
which involves n copies of $\rho_A$. In general, for a thermal bath with Hamiltonian $\hat{H}$ and temperature T, we have $\rho \sim e^{-\hat{H}/T}$ and partition function $Z = \Tr(\rho)$. Associating Euclidean time with temperature, we can then relate periods of Euclidean time to the partition function on a Riemann surface, and, more generally, on an n-sheeted Riemann surface, we have, in the ground state \cite{Calabrese2004,Hertzberg2013}
\begin{equation}
    \label{tr in terms of partition}
    \Tr(\rho_A^n) = \frac{Z_n}{Z_1^n} \Leftrightarrow \ln(\Tr(\rho^n)) = \ln(Z_n) - n \ln(Z_1)
\end{equation}
In a general QFT setting, the entanglement entropy is that of the correlations between the vacua of $A$ and $\bar{A}$ and their fluctuations, i.e. each order of the expansion in the relevant couplings contributes to the entropy via vacuum diagrams. We can then determine the entanglement entropy as an expansion in powers of the couplings as
\begin{equation}
    \label{Von Neumann expansion}
    S_E(\rho) = \sum_{k=0}^{+\infty} S_{E,k}(\rho)
\end{equation}
so that equations \eqref{VNS replica} and \eqref{tr in terms of partition} now give
\begin{equation}
    S_{E,k}(\rho) = - \frac{\partial}{\partial n} \big[\ln(Z_{n,k}) - n \ln(Z_{1,k})\big]\Big|_{n\to 1}
\end{equation}
In practice, this is the flat space limit of the geometric entropy \cite{Callan1994,VanRaamsdonk2017} in conical space with deficit angle $\delta = 2\pi(1-n)$. In coordinates $\mathbf{x} = \{r,\theta,x_\perp\}$, the conical space metric is \cite{Inomata2012}:
\begin{equation}
    \label{Conical space metric singular}
    ds^2 = -dr^2 - n^2 r^2 d\theta^2 - dx_\perp^2
\end{equation}
where $r\geq 0$, $\theta \in [0,2\pi)$, and the $x_\perp$ are the usual Cartesian coordinates on the $d_\perp$-dimensional transverse space. We will then want to expand the partition function perturbatively in powers of the couplings
\begin{equation}
    \label{Partition function expansion}
    \ln(Z_n) = \ln(Z_{n,0}) + \ln(Z_{n,1}) + ...
\end{equation}
and determine each one individually to determine the entanglement entropy of vacuum fluctuations order by order. The story here is as follows. We take the reduced density matrix of a subregion of spacetime of the vacuum state of the QFT. This exists in itself, and has a well-defined non-perturbative Von Neumann entropy. As we expand this Von Neumann entropy in powers of the couplings, what we are doing is that we are setting a scale at which we are looking at this vacuum: at tree-level we are looking at $\rho_A$ at smaller energies (larger length scales) than at 2-loops level, etc. This means that looking at loop levels of the entropy gives us how it runs with energy scales, which is the behaviour that one expects when one considers the renormalization group flow.

\section{Green's functions in Conical Space}

\label{Greens functions in Conical Space}

We consider scalar QED, that is, the theory of charged scalar particles. The corresponding action in Euclidean space in the Feynman gauge is \cite{Srednicki2007}
\begin{equation}
\label{Scalar QED action}
S[\phi,\phi^\dagger,A] = \int d^{D}\mathbf{x} \Big(-\frac{1}{4} F_{\mu \nu} F^{\mu \nu} - \frac{1}{2}(\partial_\mu A^\mu)^2 + (D_\mu \phi)^\dagger D^\mu \phi + m^2 \phi^\dagger \phi + \frac{\lambda}{4!}(\phi^\dagger \phi)^2 \Big)
\end{equation}
where the covariant derivative is $D_\mu \equiv \partial_\mu - ieA_\mu$. The partition function on the cone is
\begin{equation}
    \label{Partition function scalar qed}
    Z^{\phi,\gamma}_n = \int [\mathcal{D}\phi \mathcal{D}\phi^\dagger \mathcal{D}A] e^{- S[\phi,\phi^\dagger,A]}
\end{equation}
up to an overall normalisation factor which we ignore. In the free theory, the interaction terms vanish so that the only contributions come from the photonic and scalar propagators.

\subsection{Massive Scalar Field}

We have that the partition function at 0 temperature on the cone of a single massive scalar field with Euclidean action $S_E[\phi]$ is
\begin{equation}
    Z_n = \int [\mathcal{D}\phi] e^{-S_E[\phi]}
\end{equation}
so that
\begin{equation}
    \ln(Z^\phi_{n,0}) = -\frac{1}{2} \ln(\det[-\Delta + m^2])
\end{equation} is the partition function for the free theory \cite{Peskin2015}. Thus
\begin{equation}
    \frac{\partial}{\partial m^2} \ln(Z^\phi_{n,0}) = -\frac{1}{2}\Tr(G_n) = -\frac{1}{2}\int_n  G_n(\mathbf{x},\mathbf{x}) d^D\mathbf{x}
\end{equation}
where
\begin{equation}
    \int_n d^D \mathbf{x} \equiv \int_0^{2\pi n} d\theta \int_0^\infty r dr \int d^{d_\perp}x_\perp 
\end{equation}

In Euclidean space, the scalar propagator satisfies the equation
\begin{equation}
    (-\Delta + m^2) G_n(\mathbf{x},\mathbf{x'}) = \delta^D (\mathbf{x}-\mathbf{x'})
\end{equation}
where in flat space
\begin{equation}
    G_1(\mathbf{x}-\mathbf{x'}) = \int \frac{d^D \mathbf{p}}{(2\pi)^D} \frac{1}{\mathbf{p}^2 + m^2} e^{i\mathbf{p}.(\mathbf{x}-\mathbf{x'})}
\end{equation}
whilst in conical space this is solved by \cite{Calabrese2004}
\begin{equation}
    G_n(\mathbf{x},\mathbf{x'}) = \frac{1}{2\pi n} \int \frac{d^{d_\perp}p_\perp}{(2\pi)^{d_\perp}} \sum_{k=0}^{+\infty} d_k \int_0^{+\infty} q dq \frac{J_{k/n}(qr) J_{k/n}(qr')}{q^2+m^2+p_\perp^2} \cos(k\frac{\theta-\theta'}{n}) e^{i \mathbf{p}_\perp . (\mathbf{x}_\perp - \mathbf{x'}_\perp)} 
\end{equation}
where $d_0=1$, $d_{k\geq1}=2$ and J is the Bessel function of the first kind. Using the Euler-Maclaurin formula
\begin{equation}
    \sum_{k=0}^{+\infty} d_k F(k) = 2 \int_0^{+\infty} F(k) dk - \frac{1}{6} F'(0) - 2\sum_{j>1}^{+\infty} \frac{B_{2j}}{(2j)!} F^{(2j-1)}(0)
\end{equation}
where $B_{2j}$ are the Bernouilli numbers, we find that, in the coincidence limit $\mathbf{x'}\to\mathbf{x}$, this becomes 
\begin{equation}
    \label{Green's function scalar coincidence}
    G_n(\mathbf{x},\mathbf{x}) = G_1(0) + f_n(r)
\end{equation}
where
\begin{equation}
    \label{Function scalar coincidence}
    f_n(r) := \frac{1}{2\pi n} \frac{1-n^2}{6n} \int \frac{d^{d_\perp} p_\perp}{(2\pi)^{d_\perp}} K_0(\sqrt{m^2+p_\perp^2}r)^2 + \text{ finite}
\end{equation}
We have
\begin{equation}
    \lim_{r\to+\infty} f_n(r) = 0
\end{equation}
and the same behaviour occurs for generic $\mathbf{x}$ and $\mathbf{x'}$:
\begin{equation}
    \label{Fields vanish at infinity}
    \lim_{r\to+\infty} G_n(\mathbf{x},\mathbf{x'}) = 0
\end{equation}
as can be expected: fields and propagators vanish at infinity.

\subsection{Photon and Proca Fields}

For a Proca field - a massive spin-1 field - the partition function of the free theory is
\begin{equation}
    \label{Proca partition function}
    \ln(Z^{\gamma,(m)}_{n,0}) = -\ln(\det[-\Delta + m^2])
\end{equation}
which reduces to $\ln(Z^{\gamma,(0)}_{n,0}) = -\ln(\det[-\Delta])$ \cite{Peskin2015} in the smooth massless limit - note the factor of 2 difference with the scalar case. In Euclidean space, the Proca propagator in Feynman gauge is
\begin{equation}
    \label{Proca propagator Green's equation}
    (-\Delta + m^2) D^{\mu \nu}_{(m),1}(\mathbf{x},\mathbf{x'}) = - g^{\mu\nu} \delta^D(\mathbf{x}-\mathbf{x'})
\end{equation}
which is solved by Fourier transform in flat space as
\begin{equation}
    \label{Proca propagator flat space}
    D^{\mu\nu}_{(m),1}(\mathbf{x}-\mathbf{x'}) = \int \frac{d^D \mathbf{k}}{(2\pi)^D} \frac{-g^{\mu \nu}}{\mathbf{k}^2 + m^2} e^{i\mathbf{k}.(\mathbf{x}-\mathbf{x'})} 
\end{equation}

Since the additional degree of freedom introduced by the mass term completely decouples with the transverse degrees of freedom in the massless limit, Maxwell-Proca theory has a smooth limit into QED. Thus, we can deduce the partition function in conical space for the free photon as
\begin{equation}
    \label{Photon partition function massless limit}
    \frac{\partial}{\partial m^2} \ln(Z^{\gamma,(m)}_{n,0})\Big|_{m_\gamma\to 0} = \lim_{m_\gamma\to 0}- \int_n g_{\mu\nu} D^{\mu\nu}_{(m),n}(\mathbf{x},\mathbf{x}) d^D\mathbf{x}
\end{equation}
It must be noted that this is not necessarily true for non-abelian QFTs, where Goldstone bosons associated to the extra degrees of freedom form a non-linear sigma model. In the case of perturbative quantum gravity, this is exemplified by the van Dam-Veltman-Zakharov discontinuity \cite{vanDam1970,Zakharov1970}, where massive quantum gravity disagrees with massless quantum gravity even in the massless limit. We thus have
\begin{equation}
    \label{Green's function photon general}
    D_{(m),n}^{\mu\nu}(\mathbf{x},\mathbf{x'}) = -g^{\mu\nu} G_n(\mathbf{x},\mathbf{x'})
\end{equation}
where, as for the scalar case, we have, in the coincidence limit $\mathbf{x'}\to\mathbf{x}$
\begin{equation}
    \label{Green's function photon coincidence}
    D_{(m),n}^{\mu\nu}(\mathbf{x},\mathbf{x}) = D_{(m),1}^{\mu\nu}(0) - g^{\mu\nu} f_n(r) \end{equation}
where $f_n(r)$ is given in equation \eqref{Function scalar coincidence}. Likewise, for the photon propagator,
\begin{equation}
    D_n^{\mu\nu}(\mathbf{x},\mathbf{x}) = D_{1}^{\mu\nu}(0) - g^{\mu\nu} f_{n,m_\gamma\to 0}(r)
\end{equation}

\section{Entanglement Entropy in the Free Theory}

\label{Entanglement entropy in the free theory}

For the scalar field contribution to the entropy, we follow the work done by Hertzberg \cite{Hertzberg2013} to get
\begin{align}
\frac{\partial}{\partial m^2} \ln(\Tr(\rho_{(\phi),0}^{n})) &= \frac{\partial}{\partial m^2} \ln(Z^{\phi}_{n,0}) - n \frac{\partial}{\partial m^2} \ln(Z^{\phi}_{1,0}) \\ &= -\frac{1}{2} \Big[\int_n  G_n(\mathbf{x},\mathbf{x}) d^D \mathbf{x} - n \int  G_1(0) d^D\mathbf{x} \Big] \\ 
&= - \frac{1-n^2}{12} \int d^{d_\perp} x_\perp \int_0^\infty dr r \int \frac{d^{d_\perp}p_\perp}{(2\pi)^{d_\perp}} K_0^2(\sqrt{m^2 + p_\perp^2}r) \\
&= -\frac{1-n^2}{24n} A_\perp \int \frac{d^{d_\perp} p_\perp}{(2\pi)^{d_\perp}} \frac{1}{m^2 + p_\perp^2}
\end{align}
where $\int d^{d_\perp}x_\perp = A_\perp$ and $\int_0^\infty dy y K_0^2(y) = \frac{1}{2}$. Thus, from \eqref{VNS replica},
\begin{align}
    S^{(\phi)}_0 &= -\frac{\partial}{\partial n} \Big[- \frac{1-n^2}{24n} A_\perp \int dm^2 \int \frac{d^{d_\perp} p_\perp}{(2\pi)^{d_\perp}} \frac{1}{m^2 + p_\perp^2} \Big]\Big|_{n\to1} \\
    &= -\Big[(\frac{1}{12} + \frac{1-n^2}{24n ^2}) A_\perp \int \frac{d^{d_\perp} p_\perp}{(2\pi)^{d_\perp}} \ln(m^2 + p_\perp^2) + \text{const}\Big]\Big|_{n\to1} \\
    &= -\frac{1}{12} A_\perp \int \frac{d^{d_\perp} p_\perp}{(2\pi)^{d_\perp}} \ln(m^2 + p_\perp^2) + \text{const}
\end{align}

Likewise, for the Proca field contribution to the entropy, we use \eqref{Green's function photon coincidence} to get
\begin{equation}
    \frac{\partial}{\partial m^2} \ln(\Tr(\rho_{(\gamma,m),0}^{n})) = \frac{D(1-n^2)}{12n} A_\perp \int \frac{d^{d_\perp} p_\perp}{(2\pi)^{d_\perp}} \frac{1}{m_\gamma^2 + p_\perp^2}
\end{equation}
where we do not have a factor of $\frac{1}{2}$ for the partition function and the Proca and scalar propagators differ by a minus sign, so that the photonic contribution to the tree-level ground state entropy is obtained in the massless limit and yields
\begin{equation}
    S_{0}^\gamma = \frac{D}{6} A_\perp \int \frac{d^{d_\perp} p_\perp}{(2\pi)^{d_\perp}} \ln(p_\perp^2) + \text{const}
\end{equation}
    
Hence, in the free scalar QED theory, the leading contribution to the vacuum entropy is
\begin{equation}
    S_0^{\phi,\gamma} = \frac{A_\perp}{12} \int \frac{d^{d_\perp} p_\perp}{(2\pi)^{d_\perp}} (2D\ln(p_\perp^2) - \ln(m_\phi^2 + p_\perp^2))
\end{equation}

\section{Entanglement Entropy in Scalar QED}

\label{Entanglement entropy in scalar qed}

Let us now consider the full interacting theory of scalar QED, given by the action \eqref{Scalar QED action} and corresponding partition function \eqref{Partition function scalar qed}. We then have
\begin{equation}
    \ln(Z^{\phi,\gamma}_n) = \ln(Z^\phi_{n,0}) + \ln(Z^\gamma_{n,0}) + \ln(Z^{(a)}_{n,1}) + \ln(Z^{(b)}_{n,1}) + \ln(Z^{(c)}_{n,1}) + \mathcal{O}(\lambda^2,e^3)
\end{equation}
with
\begin{equation}
    \ln(Tr(\rho_{(1)}^n)) = \ln(Z_{n,1}) - n \ln(Z_{1,1})
\end{equation}
so that
\begin{equation}
    S_E = S_0^{\phi,\gamma} + S_1^{(a)} + S_1^{(b)} + S_1^{(c)} + \mathcal{O}(\lambda^2,e^3)
\end{equation}

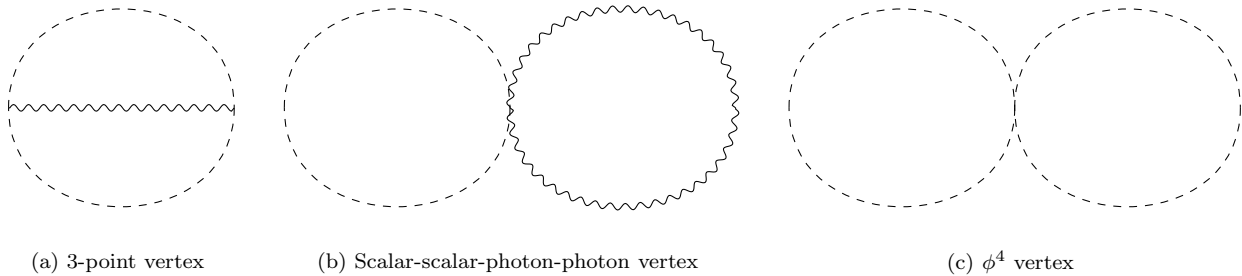
\begin{figure}[b!]
    \centering
    \subfloat[\centering 3-point vertex \label{3-point vertex vacuum diagram}]{\begin{tikzpicture}
          \begin{feynman}[scale=2,transform shape]
            \vertex (eleft);
            \vertex [right=of eleft] (eright);

            \diagram* {
                (eleft) -- [photon] (eright),
                (eleft) -- [scalar, half left] (eright),
                (eright) -- [scalar, half left] (eleft),
            };
          \end{feynman}
        \end{tikzpicture}}
        \qquad
    \subfloat[\centering Scalar-scalar-photon-photon vertex \label{4-point vertex vacuum diagram}]{\begin{tikzpicture}
          \begin{feynman}[scale=2,transform shape]
            \vertex (e);
            \vertex [right=of e](int);
            \vertex [right=of int] (p);

            \diagram* {
                (int) -- [photon,half left] (p) -- [photon,half left] (int),
                (int) -- [scalar, half left] (e) -- [scalar, half left] (int),
            };
          \end{feynman}
        \end{tikzpicture}}
    \qquad
    \subfloat[\centering $\phi^4$ vertex \label{phi4 vertex vacuum diagram}]{\begin{tikzpicture}
          \begin{feynman}[scale=2,transform shape]
            \vertex (e);
            \vertex [right=of e](int);
            \vertex [right=of int] (p);

            \diagram* {
                (int) -- [scalar,half left] (p) -- [scalar,half left] (int),
                (int) -- [scalar, half left] (e) -- [scalar, half left] (int),
            };
          \end{feynman}
        \end{tikzpicture}}        
        \caption{Vacuum contributions to the entropy from scalar QED to $\mathcal{O}(e^2)$}
\end{figure}

We start with diagram \ref{3-point vertex vacuum diagram}, where it is a priori nontrivial to apply the momentum-space Feynman rules (given in Appendix \ref{Feynman rules}) of the 3-point scalar QED interaction to position space. The diagram reads
\begin{align}
    \ln(Z^{(a)}_{n,1}) &= (-e)^2 \iint_n \expval{A^\mu(\mathbf{x})\phi(\mathbf{x})\partial_\mu \phi(\mathbf{x}) A^\nu(\mathbf{x'})\phi(\mathbf{x'})\partial'_\nu \phi(\mathbf{x'})}_0 d^D\mathbf{x} d^D\mathbf{x'} \\
    &= \frac{e^2}{4} \iint_n \expval{A^\mu(\mathbf{x})\partial_\mu( \phi^2(\mathbf{x})) A^\nu(\mathbf{x'})\partial'_\nu (\phi^2(\mathbf{x'}))}_0 d^D\mathbf{x} d^D\mathbf{x'} \\
    &= \frac{e^2}{4} \iint_n \Big(\expval{\partial_\mu(A^\mu(\mathbf{x}) \phi^2(\mathbf{x})) A^\nu(\mathbf{x'})\partial'_\nu (\phi^2(\mathbf{x'}))}_0 \\ &- \expval{\underbrace{\partial_\mu A^\mu(\mathbf{x})}_{0 \text{ by Gupta-Bleuler}}\phi^2(\mathbf{x}) A^\nu(\mathbf{x'})\partial'_\nu (\phi^2(\mathbf{x'}))}_0  \Big) d^D\mathbf{x} d^D\mathbf{x'} \\
    &= \frac{e^2}{4} \iint_n \expval{\partial_\mu(A^\mu(\mathbf{x}) \phi^2(\mathbf{x})) \partial'_\nu(A^\nu(\mathbf{x'}) \phi^2(\mathbf{x'}))}_0 d^D\mathbf{x} d^D\mathbf{x'} \\
    &= \frac{e^2}{4} \iint_n \expval{\partial_\mu \partial'_\nu (A^\mu(\mathbf{x}) \phi^2(\mathbf{x}) A^\nu(\mathbf{x'}) \phi^2(\mathbf{x'}))}_0 d^D\mathbf{x} d^D\mathbf{x'} \\
    &= \frac{e^2}{4} \iint_n \partial_\mu \partial'_\nu \expval{(A^\mu(\mathbf{x}) \phi^2(\mathbf{x}) A^\nu(\mathbf{x'}) \phi^2(\mathbf{x'}))}_0 d^D\mathbf{x} d^D\mathbf{x'} \\
    &= \frac{e^2}{2} \iint_n \partial_\mu \partial'_\nu \Big(D^{\mu\nu}_{n}(\mathbf{x},\mathbf{x'}) G_n(\mathbf{x},\mathbf{x'})^2\Big)  d^D\mathbf{x} d^D\mathbf{x'}
\end{align}

This is a total derivative of propagators which vanish at infinity as was seen in equation \eqref{Fields vanish at infinity}, so the surface term contribution will be finite and thus subleading to the order of interest at infinity. Thus, we only need to consider the coincidence limit $\mathbf{x'} \to \mathbf{x}$ for $r\to0$, i.e. the only boundary contribution at this order is that at the tip of the cone. By the divergence theorem, and since the flux is $n_\mu = \delta^r_\mu$ i.e. we go away from the tip of the cone, we have
\begin{equation}
    \iint_n \partial'_\mu d^D\mathbf{x'} d^D\mathbf{x} \to \int_0^{2\pi n} d\theta' \int_\epsilon d^{d_\perp} x'_\perp n_\mu \int_n d^D\mathbf{x} \to 2\pi n \epsilon^{d_\perp} \delta^r_\mu \int_n d^D\mathbf{x}
\end{equation}
where we integrated at coincidence, so that
\begin{align}
    \ln(Z^{(a)}_{n,1}) &= \frac{e^2 \epsilon^{d_\perp}}{2} 2 \pi n \int_n \partial_\nu (D_{n}^{\mu\nu}(\mathbf{x},\mathbf{x}) G_n(\mathbf{x},\mathbf{x})^2) \delta^r_\mu d^D \mathbf{x} \\
    &= \frac{e^2 \epsilon^{d_\perp}}{2} (2\pi n)^2 A_\perp \int_0^\infty \partial_r [(D^{r r}_1(0) - g^{r r} f_{n,m_\gamma\to 0}(r))(G_1(0)+f_n(r))^2] r dr \\
    &= \frac{e^2 \epsilon^{d_\perp}}{2} (2 \pi n)^2 A_\perp \Big[r (D^{rr}_1(0) + f_{n,m_\gamma\to 0}(r))(G_1(0)+f_n(r))^2\Big]_0^\infty \\ &- \frac{e^2 \epsilon^{d_\perp}}{2} (2 \pi n)^2 A_\perp \int_0^\infty (D^{rr}_1(0) + f_{n,m_\gamma\to 0}(r))(G_1(0)+f_n(r))^2 dr
\end{align}
as $g^{rr} = -1$. At $r\to \infty$, propagators go to 0 faster than linearly so that contribution vanishes, and note that $\lim_{r\to0} r f_n(r) \sim \lim_{r\to0} r \log^n(r) = 0$ so the first term vanishes. The non-vanishing contribution comes from the integral term. Thus,
\begin{align}
    \ln(Z^{(a)}_{n,1}) &= -\frac{e^2 \epsilon^{d_\perp}}{2} (2 \pi n)^2 A_\perp \int_0^\infty \Big[D_1^{rr}(0) G_1(0)^2 + f_{n,m_\gamma\to 0}(r) G_1(0)^2 + 2 D_1^{rr}(0) G_1(0) f_n(r) \\ &+ D_1^{rr}(0) f_n(r)^2 + 2 f_{n,m_\gamma\to 0}(r) f_n(r) G_1(0) + f_{n,m_\gamma\to 0} f_n(r)^2 \Big] dr
\end{align}
The first term is just the flat space contribution, and the contributions to the entropy of the terms with $(2\pi n)^2 f_n(r)^k$ vanish when we take the derivative with respect to n and set $n=1$ unless $k=1$. Moreover, $\frac{\partial}{\partial n}\Big[(2\pi n)^2 \frac{1}{2\pi n} \frac{1-n^2}{6n}\Big]\Big|_{n\to1} = -\frac{2\pi}{3}$. We further have that 
\begin{equation}
    \int_0^\infty K_0(k r)^2 dr = \frac{\pi^2}{4k}
\end{equation}
Thus, 
\begin{align}
    S_1^{(a)} &= -\frac{\partial}{\partial n} [\ln(Z^{(a)}_{n,1}) - n \ln(Z^{(a)}_{1,1})]\Big|_{n\to 1} \\
    &= -\frac{\pi^3 e^2 \epsilon^{d_\perp}}{12} A_\perp G_1(0) \int \frac{d^{d_\perp}p_\perp}{(2\pi)^{d_\perp}} \Big(2 \frac{G_{1,m_\gamma\to 0}(0)}{m_\phi^2+p_\perp^2} + \frac{G_1(0)}{p_\perp^2}\Big)
\end{align}
where, importantly, the area law is explicit and the overall contribution is negative. Furthermore, for diagram \ref{4-point vertex vacuum diagram},
\begin{align}
    \ln(Z^{(b)}_{n,1}) &= 2e^2 \int_n \expval{A^\mu(\mathbf{x}) A_\mu(\mathbf{x})\phi^2(\mathbf{x})}_0 d^D \mathbf{x} \\
    &= 2e^2 \int_n g_{\mu\nu} D_{n}^{\mu\nu}(\mathbf{x},\mathbf{x}) G_n(\mathbf{x},\mathbf{x}) d^D \mathbf{x} \\ &= 2e^2 \int_n g_{\mu\nu} 
 (D_1^{\mu\nu}(0) - g^{\mu\nu} f_{n,m_\gamma\to 0}(r))(G_1(0)+f_n(r)) d^D \mathbf{x} \\ 
    &= n\ln(Z^{(b)}_{1,1}) \nonumber \\ &- 2D e^2  A_\perp (2\pi n) \Big[G_{1,m_\gamma\to 0}(0) \int_0^\infty f_n(r) r dr + G_{1}(0)  \int_0^\infty f_{n,m_\gamma\to 0}(r) r dr + \int_0^\infty f_{n,m_\gamma\to 0}(r) f_n(r) r dr \Big]
\end{align}

This last term vanishes when we differentiate with respect to n and set $n=1$, and $\int_0^\infty K_0(kr) r dr = \frac{1}{2k^2}$ so that
\begin{align}
    S_1^{(b)} &= -\frac{\partial}{\partial n} [\ln(Z^{(b)}_{n,1}) - n \ln(Z^{(b)}_{1,1})]\Big|_{n\to 1} \\
    &= D e^2  A_\perp \frac{\partial}{\partial n}\Big[\frac{1-n^2}{6n} \int \frac{d^{d_\perp} p_\perp}{(2\pi)^{d_\perp}} \Big(\frac{G_{1,m_\gamma\to 0}(0)}{m_\phi^2+p_\perp^2} + \frac{G_{1}(0)}{p_\perp^2} \Big) \Big] \Big|_{n\to 1} \\
    &= -\frac{De^2}{3} A_\perp \int \frac{d^{d_\perp} p_\perp}{(2\pi)^{d_\perp}} \Big(\frac{G_{1,m_\gamma\to 0}(0)}{m_\phi^2+p_\perp^2} + \frac{G_{1}(0)}{p_\perp^2} \Big)
\end{align}
For the $\phi^4$ 2-loop diagram \ref{phi4 vertex vacuum diagram}, we have \cite{Hertzberg2013}
\begin{align}
    \ln(Z^{(c)}_{n,1}) &= - \frac{\lambda}{4!} \int_n \expval{\phi(\mathbf{x})^4}_0 d^Dx \nonumber \\
    &= - \frac{3\lambda}{4!} \int_n G_n(\mathbf{x},\mathbf{x})^2 d^Dx 
\end{align}
where the factor of 3 comes from Wick's theorem. We have $G_n(\mathbf{x},\mathbf{x})^2 = G_1(\mathbf{x}-\mathbf{x'}) + 2G_1(0) f_n(r) + f_n(r)^2$, so that
\begin{align}
    \ln(Z^{(c)}_{n,1}) - n \ln(Z^{(c)}_{1,1}) &= -\frac{3\lambda}{4!} \Big[\int_n G_n(\mathbf{x},\mathbf{x})^2 d^Dx - n \int_n G_1(\mathbf{x},\mathbf{x})^2 d^Dx \Big] \\
    &= -\frac{3\lambda}{4!} A_\perp 2\pi n \int_0^{+\infty} [2G_1(0) f_n(r) + f_n(r)^2]r dr \\
    &= -\frac{3\lambda}{4!} A_\perp \Big[\frac{1-n^2}{6n} G_1(0) \int \frac{d^{d_\perp}p_\perp}{(2\pi)^{d_\perp}} \frac{1}{m_\phi^2 + p_\perp^2} + 2\pi n \int_0^{+\infty} f_n(r)^2 r dr\Big]
\end{align}
The second term vanishes when we differentiate with respect to n and set n=1. Thus, the $\phi^4$ 2-loop contribution to the entropy is
\begin{equation}
    S_1^{(c)} = -\frac{\lambda}{4!} A_\perp G_1(0) \int \frac{d^{d_\perp} p_\perp}{(2\pi)^d_{\perp}} \frac{1}{m_\phi^2 + p_\perp^2} 
\end{equation}

Hence, putting everything together, we have
\begin{align}
    S_E &= A_\perp \int \frac{d^{d_\perp} p_\perp}{(2\pi)^{d_\perp}} \Big[\frac{1}{12}  (2D\ln(p_\perp^2) - \ln(m_\phi^2 + p_\perp^2)) \nonumber \\ &- \Big(\frac{\pi^3 e^2 \epsilon^{d_\perp}}{12} G_1(0) \big(2 \frac{G_{1,m_\gamma\to 0}(0)}{m_\phi^2+p_\perp^2} + \frac{G_1(0)}{p_\perp^2}\big)  + \frac{De^2}{3}\big(\frac{G_{1,m_\gamma\to 0}(0)}{m_\phi^2+p_\perp^2} + \frac{G_{1}(0)}{p_\perp^2}\big) + \frac{\lambda}{4!} \frac{G_1(0)}{m_\phi^2 + p_\perp^2} \Big)  \Big] + ...
\end{align}
where the area law is explicit at every order, as expected. From this, one may straightforwardly deduce the $\epsilon$ divergences. In particular, in $D=4$ dimensions ($d=3,d_\perp=2$), the momentum integral leads to a (subleading) logarithmic divergence, and the Green's functions at coincidence are quadratically divergent in momentum $\sim \Lambda^2$ where $\Lambda$ is a momentum cutoff, so regulating in position space we get $G_1(0) \sim \frac{1}{\epsilon^2}$, with $S\sim A ((e^2+\lambda)/\epsilon^2)$. Importantly, we see that entanglement entropy is cutoff-dependent, i.e. it depends on the details of the microphysics. In particular, putting this theory on a lattice would make entanglement entropy dependent on lattice spacing and configurations.

\bigskip

Furthermore, we see that these loop contributions necessarily reduce the entropy. How do we make sense of this in light of the flow of the couplings? Indeed, to leading order in $\lambda$ and $e$, we have \cite{Srednicki2007}
\begin{align}
    \beta_e(e,\lambda) &= \frac{e^3}{48\pi^2} \label{beta function e scalar qed}\\
    \beta_\lambda(e,\lambda) &= \frac{1}{16\pi^2} (5\lambda^2 - 16 \lambda e^2 + 24 e^4) \label{beta function lambda scalar qed}
\end{align}
Both $\beta$ functions are positive, so couplings increase with energy (and so decrease with distance). In practice, this means that the 2-point correlation functions of scalar QED ought to increase with energy, i.e. as we ``zoom in" to smaller and smaller distances. In particular, this can be done at the boundary between $A$ and $\bar{A}$, i.e. the correlations between both regions increase as we ``zoom in" - we expect this to diverge at the Landau pole. However, the Von Neumann entropy of region A - i.e. the entanglement entropy - seems to decrease as we ``zoom in". On the one hand, we seem to have that correlation between both regions increase at the boundary; on the other hand, the reduced density matrix $\rho_A$ becomes more and more pure, so entanglement between the two regions shrinks. How do we solve this apparent contradiction?

\bigskip

There are several ways to see this. The particle physics way is to consider that at higher energies, the contributions to the local physics are getting increasingly local, i.e. long-distance contributions are subleading to local fluctuations - this is reminiscent of the concept of screening in QED. As one probes the vacuum to higher and higher energies (smaller and smaller distances), one gets more and more vacuum fluctuations, so the physics will look more and more local. At the Landau pole, the local physics completely dominates, and $A$ is entirely independent of $\bar{A}$ with no entanglement between the two regions. 

\bigskip

Thus, the local vacuum purifies with increasing energy. From the point of view of quantum information theory, this is in agreement with the general trait of entanglement ``monogamy'' \cite{Seevinck2010,Coffman2000} - whereby a subsystem strongly interacting and thus entangled with a second subsystem cannot be strongly entangled with a third one. 

\section{Conclusions and outlook}

For the usual spinor QED, we expect to see the same behaviour. For perturbative (asymptotically safe) quantum gravity, we also expect this. For Quantum ChromoDynamics (QCD), however, it would be interesting to explore how asymptotic freedom and the negative beta functions influence the flow of entanglement entropy and whether antiscreening has the opposite effect on entropy at high energies. We indeed conjecture that the beta function and RG flow of couplings dictate the flow of the entanglement entropy. Thus, we expect the entanglement entropy to be scale invariant in supersymmetric theories where loop contributions cancel out precisely and beta functions are zero.

\bigskip

This work may also be straightforwardly expanded to consider entropy and entanglement renormalization within NLQFT, as has recently been done for $\phi^4$ theory \cite{Landry2024}, where entanglement entropy was shown to be free of UV divergences. This result is expected to carry on to scalar QED. 

\bigskip

The methods and results obtained may also be contrasted to those obtained in condensed matter systems. Indeed, there has recently been a lot of interest in understanding the area law of entanglement entropy in such contexts - e.g. see \cite{Hastings2007,Laflorencie2016} and references therein. For instance, (spinor) QED in $D=1+2$ dimensions is related to the spin-1/2 Heisenberg antiferromagnet, and the RG flow of couplings in such a discretised system has been studied \cite{Thomson2017}, where it was found that for generic disorder the flow led to strong couplings. It may thus be insightful to analyse the flow of entanglement entropy in such systems, which may be readily compared with our results by regularising the quantum field theory, and see whether entanglement entropy also follows an opposite behaviour to that of couplings under the renormalization group.

\bigskip

This paper constitutes an early investigation of entangling behaviours in somewhat realistic QFTs, which may be extended to non-abelian theories such as QCD and perturbative quantum gravity. However, in such cases, deriving the partition function from Green's function might be more tricky as Green's functions of the massive non-abelian theories do not reduce to those of the massless theories in the limit $m\to0$. This becomes an issue because there is then nothing to differentiate with respect to in, say, equation \eqref{Proca partition function}. Thus, the tree-level (leading) contributions might not be determined using this approach, but the flow of entanglement entropy is independent of these considerations and may be examined. This interesting problem will be tackled in a future paper. 

\section{Acknowledgements}

S.F. is funded by a studentship from the Engineering and Physical Sciences Research Council and thanks Adrian Kent for his continuous support. SB thanks EPSRC grants EP/R029075/1 and EP/X009467/1.

\appendix

\section{Feynman Rules of Scalar QED}

\label{Feynman rules}

Here we recall the Feynman rules of interest in Fourier space for scalar QED in Euclidean spacetime :
\begin{itemize}
    \item For each scalar-scalar-photon vertex, write $-e(k+k')_\mu$
    \item For each scalar-scalar-photon-photon vertex, write $2e^2 g_{\mu\nu}$
    \item For each $\phi^4$ vertex, write $-\lambda$
    \item For each internal photon, write $\frac{-g^{\mu\nu}}{k^2}$ (in Feynman gauge)
    \item For each internal scalar, write $\frac{1}{k^2 + m^2}$
\end{itemize}

\bibliographystyle{apsrev}
\bibliography{library}

\end{document}